\documentclass[%
preprint,
 amsmath,amssymb,
 aps,
]{revtex4-2}

\usepackage{graphicx}
\usepackage{dcolumn}
\usepackage{bm}
\usepackage[usenames,dvipsnames,svgnames,table,rgb]{xcolor}

\usepackage{array,tabularx,tabulary,booktabs} 
\usepackage{longtable}  
\usepackage{multirow} %

\begin{document}


\title{Branching random graph model of rough surfaces describes thermal properties of the effective molecular potential}





\author{Aleksey Khlyupin}
 \email{khlyupin@phystech.edu}
 \affiliation{Moscow Institute of Physics and Technology, Institutskiy Pereulok 9, Dolgoprudny, Moscow 141700, Russia}
 
\author{Timur Aslyamov}
 \email{t.aslyamov@skoltech.ru}
\affiliation{ Center for Design, Manufacturing and Materials,
   Skolkovo Institute of Science and Technology,
  Bolshoy Boulevard 30, bld. 1, Moscow, Russia 121205
}%



\begin{abstract}
Fluid properties near rough surfaces are crucial in describing fundamental surface phenomena and modern industrial material design implementations. One of the most powerful approaches to model real rough materials is based on the surface representation in terms of random geometry. Understanding the influence of random solid geometry on the low-temperature fluid thermodynamics is a cutting edge problem. Therefore this work extends recent studies bypassing high-temperature expansion and small heterogeneity scale. We introduce random branching trees whose topology reflects the hierarchical properties of a random solid geometry. This mathematical representation allows us to obtain averaged free energy using a statistical model of virtual clusters interacting through random ultrametric pairwise potentials. 
Our results demonstrate that a significant impact to fluid-solid interface energy is induced by the hierarchical structure of random geometry at low temperature. These calculations coincide with direct Monte Carlo simulations. Due to the study's interdisciplinary nature, the developed approach can be applied to a wide range of quenched disorder systems on random graphs. 
\end{abstract}

\maketitle


\section{\label{sec:intro} Introduction}

The study of molecular fluid thermodynamic properties near solid surfaces with nanoscale geometric heterogeneity is an urgent problem in modern physics. Although thousands of works are devoted to studying confined molecular systems, most of them consider the solid surface model to be smooth on a nanometer scale. However, in recent years, more and more publications demonstrate the significant effect of nanoroughness on a broad class of physical and chemical phenomena: adsorption of simple and chain fluids for materials characterization  \cite{neimark2009quenched, ravikovitch2006density, aslyamov2017density, aslyamov2019random, hlushak2018heat, landers2013density, jagiello2013carbon} and optimal storage design \cite{aslyamov2019theoretical}, thin-film thermodynamics and wetting transition on atomic-scale roughness substrates \cite{khlyupin2016effects, zhou2018wetting, katasho2015mechanisms, herminghaus2012universal, yatsyshin2017classical}, friction properties of superhydrophobic surface for liquid lubrication design \cite{hu2019molecular}, molecular surface modifiers for supercapacitors \cite{fileti2019investigating}, heterogeneous catalysis \cite{nartova2015model}, thermal conductivity at interfaces and confinement \cite{surblys2019molecular, coasne2013adsorption} and even geometric control of bacterial surface accumulation \cite{mok2019geometric}.

In computer simulations, geometrical heterogeneity corresponds to the corrugated zones of molecular layers or surface chemical groups. Therefore, to take into account surface roughness, the solid-fluid interaction potential explicitly depends on the geometrical defects' locations. Such full positional detalization is numerically expensive and also overcomplicated to fit the experimental data. The significant increase of computational speed without serious prejudice to the accuracy can be obtained using the effective solid-fluid potential to reduce the geometrical heterogeneity detalization \cite{shi2019bottom}. Besides surface geometry, one of the most crucial properties of the effective potentials is temperature dependence. The appropriate approach was developed in work \cite{forte2014effective} as a free-energy averaged (FEA) mapping of an exact system to a coarse-grained model. However, even coarse-grained models require serious computational power. Despite the permanent growth of computer technologies, numerical costs can only be reduced in insignificantly rough materials. Indeed the geometrical model of significantly rough materials demands to consider a substantial number of the corrugated solid molecular layers. Thus, the characterization of the materials exhibiting the roughness region around 15-20\AA (meso-porous silica) using direct simulations methods is impossible.  

One of the most realistic model of the significantly rough surfaces is the realisation of the correlated random processes. In our work \cite{khlyupin2017random}, we developed the effective potential of the interaction between a fluid molecule and the correlated random surface. We calculated free-energy-averaged (FEA) effective interaction potentials using an analogy with the First Passage Time Probability Theory for Markov random processes.
The resulting potential is used as an input part of density functional theory \cite{aslyamov2017density, aslyamov2019random}, which allows us to describe fluid properties near significantly rough materials \cite{aslyamov2019theoretical}. However, the base model is limited to the leading-order term in the high-temperature expansion of the FEA potential, which is inaccurate at low temperatures. Recently Shi, Santiso and Gubbins in \cite{shi2019bottom} pointed out the limitations of our model and emphasized the need for the potential that would work well for surfaces with large geometrical defects. Also, more accurate temperature dependence is needed for calculations of isosteric heat of adsorption, where the influence of surface heterogeneity is crucial \cite{cimino2017determination}.

\begin{figure}[b]
    \centering
    \includegraphics[width=\textwidth]{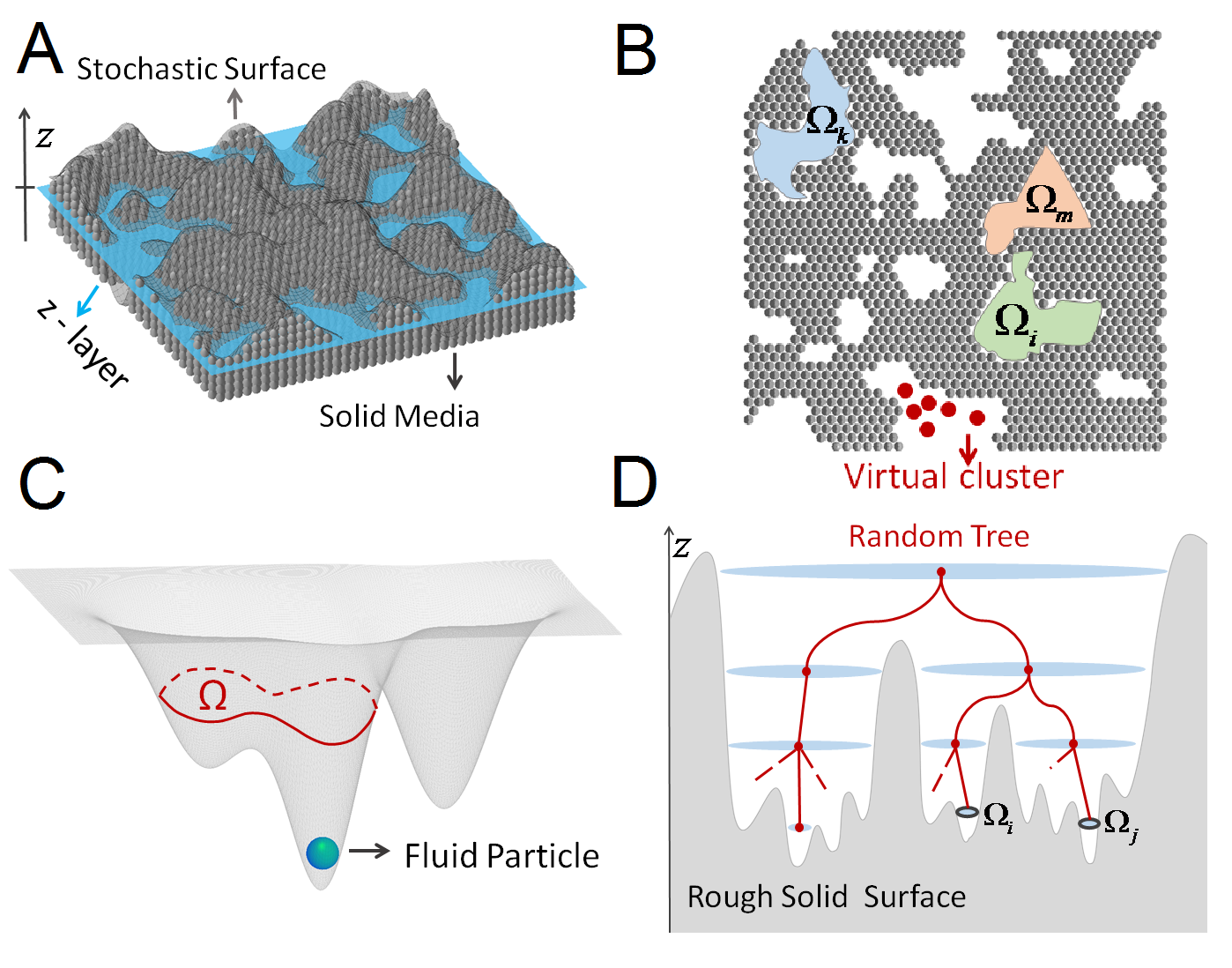}
    \caption{Illustration of 3D solid surface geometry with a slice at level z. Gray balls illustrate solid atoms. Solid surface can be described by certain realization of 
the 2D random field (A). Configuration space $\Omega_T=\cup_i\Omega_i$ is available for fluid molecule at $z$-slice (B). Colors reflect fluctuations of the external field in each closed region ("islands") $\Omega_i$. Excluded volume that determines the interaction of a fluid particle with a solid media (C). Illustration of a hierarchical surface of a solid phase and a branching random graph that reflects the structural features of such random geometry (D).}
    \label{fig:Fig1}
\end{figure}

Here we develop a statistical physics model to expand the FEA fluid-solid interaction potential calculations in the field of low temperatures and a wide range of geometric heterogeneity scales.  We study an object well defined from a physical point of view — a configuration integral for a single fluid molecule, taking into account the interaction with atoms of the solid media. However, unlike works in which this integral is calculated numerically using Monte Carlo simulations \cite{forte2014effective,shi2019bottom}, we propose a theoretical approach introducing an ensemble of branching graphs that reflect the hierarchical structure of random solid geometry (see Fig \ref{fig:Fig1}). The following sections are devoted to the technical details of calculating this kind of one-particle free energy. 

The article is organized as follows: a description and derivation of the probabilistic model for the joint distribution of fields is given in Section II. This section also shows how the proposed probabilistic model calculates the desired thermodynamic properties via the replica technique. Section II.B is devoted to the solution for the partition function of our quenched disorder system. A nonlinear integral equation is obtained for the probability density of effective local fields, which determines the system's free energy in the framework of a replica-symmetric ansatz. We applied the population dynamics algorithms to carry out a numerical analysis of these equations. A discussion of numerical results is given in Section III. Also, in this section, the results of comparing the obtained model with numerical simulation by the Monte Carlo method are considered; In conclusion, we discuss the applications and future development of our statistical model on branching random graphs both in problems of fluid physics and in contemporary topical bioinformatics problems associated with random graphs, problems of cooperation in stochastic complex social networks, swarms, and populations. 

\section{\label{sec:model} Model}

From the physics point of view, the roughness is a relative characteristic, and the size of a fluid particle defines the influence of the geometry. For example, a particle with a much larger size than a roughness scale does not interact with heterogeneity and the surface can be considered smooth. The more exciting case corresponds to molecules much smaller than the roughness that can penetrate the solid matter's free space. The central object of our research is a probe molecule located within significantly rough solid media. We consider a molecule of fluid interacting with a solid phase, consisting of $ M $ atoms located at the fixed sites of some three-dimensional lattice. The solid media is bounded from above by the surface, which corresponds to random process realization $Z(\vec{r})$. Without losing generality, we assume the mean value of $Z$ equals zero. Figure~\ref{fig:Fig1}A also illustrates the available fluid space, which is free of the solid molecules. Let us consider a probe molecule at the distance $z$. The partition function of the single-molecule interacting with solid media is defined by the integration over the configuration space $\Omega_T(z)$ available at the $z$ level:

\begin{equation}
    \Theta=\int_{\Omega_T}d\vec{r}e^{-\beta U(\vec{r})}
\end{equation}

\noindent where $U(\vec{r})$ is the total potential of the interaction between fluid molecule and all solid atoms located at fixed points $\{\vec{r}_{s,k}\}_{k=1}^M$. The total potential can be represented as sum of pairwise additive Lenndrd-Jones (LJ) interactions $U(\vec{r})=\sum_{k=1}^M{U_{LJ}(\vec{r},\vec{r}_{s,k})}$. We introduce the symbol $\beta = 1/k_BT$, where $T$ is the temperature and $k_B$ is Boltzmann constant. Fig.~\ref{fig:Fig1}B illustrates the slice of the solid media at the level $z$, where the total configuration space can be split into separated regions ("islands") $\Omega_T=\cup_i\Omega_i$ with different mean potentials $U_i$ at different regions. Thus, the partition function can be represented as the sum:
\begin{equation}
\label{eq:pf_exact}
    \Theta=\sum_i\int_{\Omega_i}d\vec{r}e^{-\beta U_i(\vec{r})}
\end{equation}
For the sake of simplicity we assume that all the islands have the same average size $|\Omega_i|=|\Omega|_z$ depending on the slice level $z$. Wherein the random surface properties are conserved and taken into account by the random values of external potentials $U_i$ corresponding to the island $\Omega_i$. The Free Energy Averaging procedure \cite{forte2014effective, khlyupin2017random} allows us obtain the effective 1D fluid-solid interaction potential at each slice z (which depends only on the height $z$ normal to the surface). This approach contains two consisten steps: the first step is the calculation if the free energy which depends on certain realization of quenched random fields $U_i$; the second step is the averaging using joint probability distribution $P(U_1,U_2,...,U_N)$ of random fields as:
\begin{eqnarray}
\label{eq:LogF}
   -\beta F=\langle \log \Theta \rangle_P= \int dU_1...dU_N P(U_1,...,U_N)\log \Theta(U_1,...,U_N)
\end{eqnarray}
where $F$ is the Helmholtz free energy of the probe particle. Thus, physical properties of the system are mostly defined by random fluctuations of fields $U_i$ and their correlations. Therefore, a deep analysis of their joint probability distribution is required.

We would like to briefly demonstrate a sketch of the proposed idea from a bird's eye view. To build a theoretical model avoiding the high-temperature approximation we introduced random branching trees whose topology reflects the hierarchical properties of random solid geometry with large scale of heterogeneity (see Fig~\ref{fig:Fig1}.D). The joint probability model $P(U_1,...,U_N)$ describes the correlations between random fields taking into account such hierarchical structure of rough solids. To meet these requirements, we introduce the simple model of summation of random variables on the graph branches. Proposed probabilistic model of random fields on random branching trees allows to link the calculation of single particle averaged free energy with the statistical model of virtual clusters interacting through random pairwise potentials. The problem of calculating the free energy averaged over the random fields or random pair interactions naturally arises in the theory of systems with “frozen” or quenched disorder: spin glass theory \cite{mezard1987spin,schneider1977random, sherrington1975solvable, mezard2001bethe}, theory of spin systems on random graphs \cite{nikoletopoulos2004replicated, erichsen2011phase, erichsen2017multicritical}, spectral theory of random matrices and graph Laplacian \cite{kuhn2008spectra,rogers2008cavity,rogers2010spectral, dean2002approximation}. Thus to solve proposed model of virtual clusters, we developed a method based on the finite-connectivity technique for calculating the partition function of spin systems in random small-world networks and applied this method to study fluid behavior for the first time.




\subsection{Probability model for Random Fields on Branching Random Graphs}

Let us consider the field $U_i$, which corresponds to the Lennard Jonnes interaction of a particle located in the island $\Omega_i$ in the solid media. Considering the space above and below the particle at level $z$ the islands form the volume $\Sigma_i=\int dz\Omega_i(z)$, where integration corresponds to the normal direction (Fig~\ref{fig:Fig1}C). Then the solid-fluid interaction $U_i$ can be found as the integral over the solid media $V_s = V_{tot}\setminus \Sigma_i$  - total volume $V_{tot}$ without volume $\Sigma_i$.
\begin{equation}
\label{eq:Ui_integral}
    U_i=\rho_s\int_{V_s} d\vec{r_s} U_{LJ}(|\vec{r}_i-\vec{r}_s|)
\end{equation}

\noindent where  $U_{LJ}$ is the Lennard-Jonnes potential between the particle and solid molecules and $\rho_s$ - number density of the solid media. Due to the rapidly decreasing Lennard-Jonnes potential, the contribution from each layer to (\ref{eq:Ui_integral}) is mainly determined by the geometric properties of the islands  $\Omega_i(z)$ (the size $|\Omega|_z$ and shape), which are random parameters and vary for each realization of a rough solid geometry. Thus, the external field in each $\Omega_i(z)$ at level z is determined by all solid media layers and the islands in such the hierarchical structure Fig~\ref{fig:Fig1}D. This hierarchical structure of the rough surfaces imposes certain requirements on the fluctuations of the fields $U_i$. Thus, the joint probabilistic model $P(U_1,...,U_N)$ describes the correlations between random fields fluctuations taking into account the hierarchical structure of the islands in normal z-direction.

\begin{figure}
    \centering
    \includegraphics[width=\textwidth]{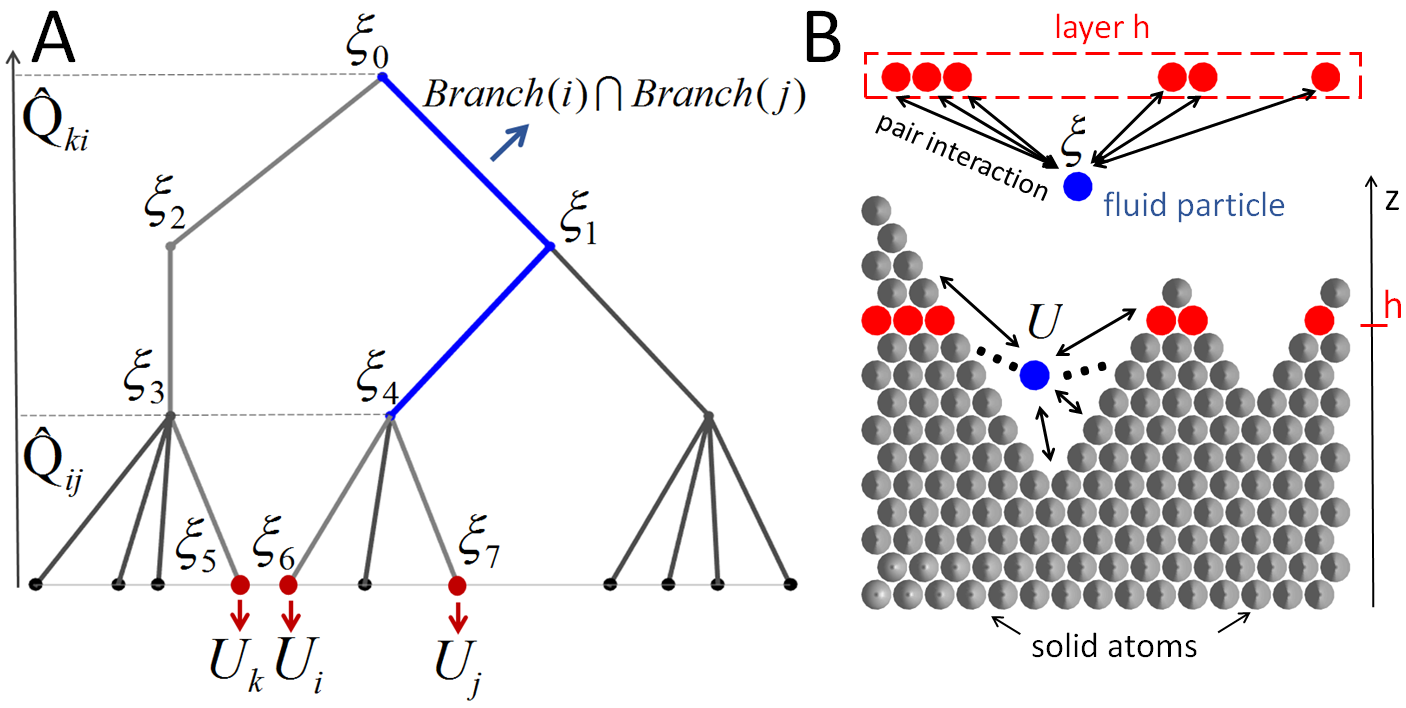}
    \caption{Schematic illustration of a branching random graph and the process of summing random variables $\xi_i$ along branches to generate random external fields $U_k$. The blue solid line indicates a common path in the graph from the root to leafs $i$ and $j$ (A). Schematic illustration of physical interaction of fluid molecule with all solid atoms  - $U$ and with solid atoms from fixed layer - $\xi$ (B).}
    \label{fig:Fig2}
\end{figure}

In our model each realization of a random geometry of solid media with such hierarchical structure of $\Omega_i(z)$ corresponds to the Branching Random Tree (BRT), where the vertices reflect islands $\Omega_i$ Fig.~\ref{fig:Fig1}D.  
Let us consider the layer $z$ of the solid media and a set of islands $\Omega_i(z)$. The contribution to the fields $U_i$ from the layer at level $h$ depends on the random geometry of the islands $\Omega_i(h)$. Therefore, the contributions of vertices $i$ in the graph BRT Fig.~\ref{fig:Fig2}A  are the random variable $\xi_i$ which reflects the impact ot the corresponding layer into solid media Fig.~\ref{fig:Fig2}B. We assume that all $\xi_i$ are independent random variables with a certain expectations and a variances. One can replace the integration over the layers in (\ref{eq:Ui_integral}) by summation over the vertices in RBT, which contribute to the external field $U_i$. Thus, in our model, for a fixed realization of a random graph, field  $U_i$ is defined as the sum of the random variables $\xi_k$ on the vertices of the graph that lie on the branch connecting the root of the tree and the final vertex $i$.

\begin{equation}
\label{eq:Ui_graph}
U_i=\sum_{k\in Br(i)}\xi_k 
\end{equation}
where branch $Br(i)$ is unique due to tree structure of the random graph. Figure \ref{fig:Fig10} demonstrates schematic illustration of several different realizations of hierarchical solid surface and a branching graphs (trees) that reflect the structural features of such geometry. 

\begin{figure}
    \centering
    \includegraphics[width=\textwidth]{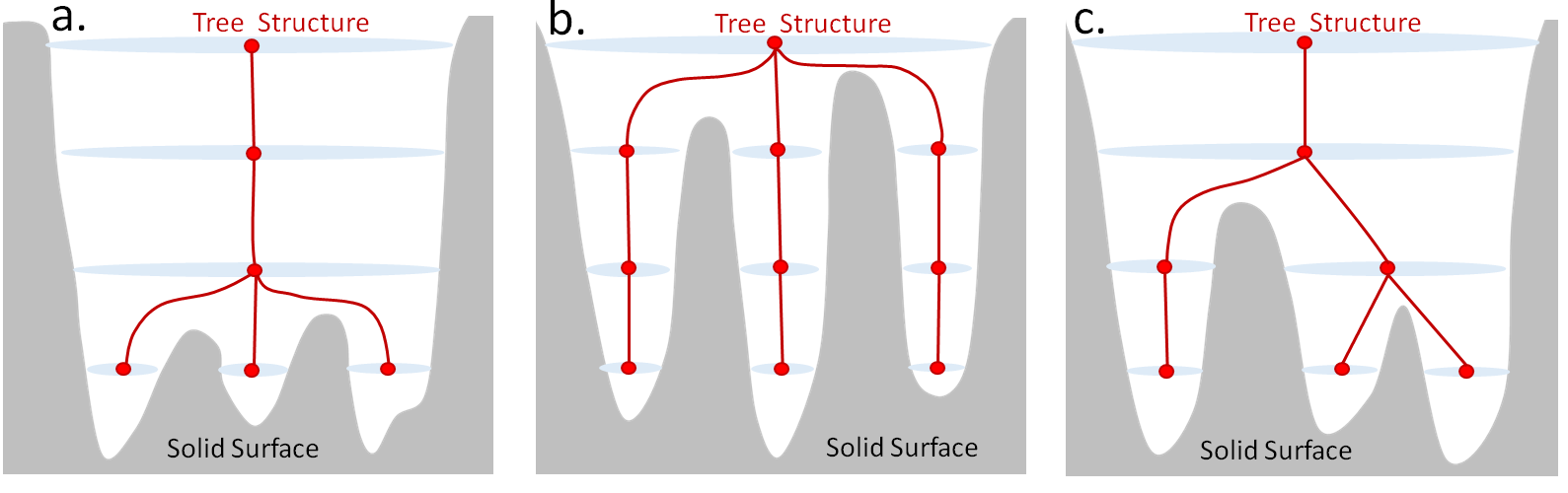}
    \caption{Schematic illustration of several different realizations of hierarchical solid surface and a branching graphs (trees) that reflect the structural features of such geometry. All these structures ultimately give rise to 3 pits inside the solid (3 red dots at the end vertices of the graph) in different ways.}
    \label{fig:Fig3}
\end{figure}

The object of the most interest is the graphs with large number of the layers. In this case the joint probability distribution of the fields $U_i$ can be described by the multivariate Gaussian model. Without loss of generality, one can consider normalized random fields  $U_i=U_i-U_0$, where $U_0=E(U_i)$ is the expected value. The crucial part of the model is the correlation matrix $C_{ij}$, which determines the fluctuations of the fields. Assuming pairwise independence of random variables $\xi_k$ correlations can be written as:
\begin{equation}
\label{eq:correlation}
    C_{ij}=E\Big[ U_i U_j\Big]=E\sum_{k\in Br(i)}\sum_{s\in Br(j)}\xi_k\xi_s=\sum_{k\in Br(i)\cap Br(j)} E\xi_k^2
\end{equation}
As one can see from (\ref{eq:correlation}) the correlation between two fields $U_i$ and $U_j$ in islands $\Omega_i$ and $\Omega_j$ is defined not by the Euclidean distance between them $|\vec{r}_i-\vec{r}_j|$, but by the common path in the Branching Tree for branches $Br(i)\cap Br(j)$. In this case the distance between the vertices induces the ultrametric space with unusual for Euclidean space properties \cite{mezard1987spin}.

We introduce the intersection $Q_{ij}$ which denotes the depth in the graph where branches $Br(i)$ and $Br(j)$ start to diverge. Using this definition one can rewrite expression (\ref{eq:correlation}) as the follows: $ C_{ij}=C(Q_{ij})$. For a specific realization of a random graph with fixed intersections $Q_{ij}$, the conditional probability of random fields $U_i$ has the form of multivariate normal distribution: 
\begin{equation}
\label{eq:cond_prob}
P(U_1,...,U_N|Q_{ij},...,Q_{kl})=\frac{1}{A}\exp \left\{-\frac{1}{2}\sum_{ij}U_iC^{-1}(Q_{ij})U_j\right\}
\end{equation}
where $A$ is the normalization constant. Each realization of a random geometry of the solid surface has its own realization of random graph. Therefore, to obtain the final probability measure for fields $U_i$, it is necessary to average the conditional probability (\ref{eq:cond_prob}) over an ensemble of random graphs. This ensemble defines the joint probability of all intersect ions $P(Q_{ij},...,Q_{kl})$. We assume the simple model of this joint probability $P=\prod\rho(Q_{ij})$, where each $Q_{ij}$ is chosen according to the same discrete multimodal distribution $\rho(Q_{ij})$:
\begin{equation}
    \rho(Q_{ij})=\sum_{k}p_k\delta(Q_{ij}-q_k)
\end{equation}

\noindent where discrete values $p_k$ denote the probabilities that an arbitrary pair of branches $i,j$ has intersections  $Q_{ij}$ at depth level $q_k$.
These probabilities are ($\sum p_k=1$) and $p_k$ could be calculated for each graph ensemble as relative number of pairs which satisfy the required property:
\begin{equation}
\label{eq:pk}
    p_k=\frac{\#\{(i,j):Q_{ij}=q_k\}}{N^2}=\frac{a_k}{N}
\end{equation}

\noindent The numbers (\ref{eq:pk}) depend on random graph generation patterns and can be calculated numerically by the Monte Carlo simulations or analytically using the Parisi matrix representation of such graph structure \cite{mezard1987spin}. 
Summing all the above together, the desired general probability model for the distribution of random fields has the following form:
\begin{equation}
    \label{eq:general_prob}
    P(U_1,...,U_N)=\frac{1}{A}\int \prod_{ij}dQ_{ij}\rho(Q_{ij})\exp(-\frac{1}{2}\sum_{ij}U_iC^{-1}(Q_{ij})U_j)
\end{equation}
\noindent
Obtained probability distribution for fields $U_i$ allows us to consider averaged over random surface geometry partition function as  $\langle\Theta\rangle_P$.

According to equation (\ref{eq:LogF}) direct calculation is complicated by the presence of logarithm in the subintergal expression. For this reason we use the modified replica technique as a useful tool for quenched disorder systems \cite{mezard1987spin}. Expression (\ref{eq:pf_exact}) for partition function can be rewritten in the following form: $ \Theta=\sum_i q_i$, where auxiliary variable $q_i=e^{-\beta U_i}$ is introduced for convenience. To calculate the free energy $\beta F=-\log \langle\Theta\rangle_P$  averaged over the probability (\ref{eq:general_prob}) we start from the replica trick as follows
\begin{equation}
\label{eq:F_replica}
   - \beta F=\lim_{m\to 0}\frac{\partial \langle\Theta(m)\rangle_P}{\partial m}
\end{equation}
where we introduce the partition function of  $m$ identical replicas of the system. Using multinomial formula one can obtain the following expression for replicated partition function:
\begin{equation}
\Theta(m)=\Big(\sum_i^Nq_i\Big)^m=\sum_{\phi_1...\phi_N}m!\prod_{i=1}^N\frac{q_i^{\phi_i}}{\phi_i!}\delta(m-\sum_{i=1}^N\phi_i)
\end{equation}

To obtain the replica limit (\ref{eq:F_replica}), one can to apply the inverse Z-transform to $\Xi(z)=\sum_{k=0}^{\infty}\Theta(k)z^k$ and then take
the limit according to (\ref{eq:F_replica}). However, it is convenient to obtain an expression for the desired
average free energy immediately in terms of $\Xi$, avoiding calculations of the
inverse Z-transform:

\begin{equation}
\label{eq:F_Z-transform}
   - \beta F=\frac{1}{2\pi i}\lim_{m\to0}\frac{\partial}{\partial m}\oint \langle\Xi(z)\rangle_P z^{m-1}dz=\frac{1}{2\pi i}\oint \frac{\log z}{z}\langle\Xi(z)\rangle_P dz
\end{equation}
\noindent where the integration is performed along a counterclockwise closed path encircling the origin and entirely in the region of convergence.

In our study we demonstrate that $\langle\Xi(z)\rangle_P$ can be represented in the integral form using Gamma function representation for $m!$ as follows:
\begin{equation}
\label{eq:Xi}
    \langle\Xi(z)\rangle_P=\int_0^\infty dt e^{-t} Z(t,z)
\end{equation}
\begin{equation}
     Z(t,z)=\sum_{\phi_i}\prod_{i=1}^N\frac{1}{\phi_i!}\int\frac{\prod_{ij}dQ_{ij}\rho(Q_{ij})}{A}\int\prod_i dU_i \exp (-\beta \sum_i\phi_i(U_i+H_0)-1/2\sum_{ij}U_iC^{-1}(Q_{ij})U_j) \nonumber
\end{equation}

\noindent where the sub-integral expression on the right hand side looks similar to partition function of non-interacting clusters $N=\sum_i^N\phi_i$ under the deterministic external field $H_0=U_0-\frac{1}{\beta}\log \frac{t}{z}$ and the random fields $U_i$ \cite{hill1994thermodynamics, bugaev2007exactly, aslyamov2014complex, aslyamov2014some}. The next steps are devoted to free energy calculations according to the following order: we carry out averaging over random fields $U_i$ then over random graph typologies $Q_{ij}$, next we sum over "virtual" clusters $\phi_i$, after that we integrate over auxiliary variable  $t$ and finally obtain the replica limit using inverse $Z$-transform. 
In order to simplify (\ref{eq:Xi}) one can use well known expression for multinomial Gaussian integral:
\begin{equation}
    \label{eq:Gaussian}
    \frac{1}{A}\int\prod_idU_i\exp(-1/2\sum_{ij}U_iC_{ij}^{-1}U_j+\sum_{i}U_i)=\exp (1/2\sum_{ij}\phi_i C_{ij}\phi_j)
\end{equation}
\noindent Substituting expression (\ref{eq:Gaussian}) one can rewrite (\ref{eq:Xi}) as follows:
\begin{equation}
\label{eq:Xi_2}
Z(t,z)=\Big\langle\sum_{\phi_1...\phi_N}\prod_{i=1}^N\frac{1}{\phi_i!}e^{-\beta H} \Big \rangle_{\{Q_{ij}\}}
\end{equation}
\begin{equation}
\label{eq:Hamiltonian}
H=H_0\sum_i\phi_i-\beta\sigma^2/2\sum_i\phi_i^2-\beta/2\sum_{i\neq j}\phi_i C_{ij}(Q_{ij})\phi_j
\end{equation}
Where the replicated partition function (\ref{eq:Xi_2}) is expressed in terms of the effective Hamiltonian $H$ of interacting cluster systems with self-action $\sigma^2$ which stands for diagonal elements of the correlation matrix $C_{ij}$. The couplings $C_{ij}\in\mathbb{R}$ are independent identically distributed random variables as a decreasing functions of ultrametric distance $Q_{ij}$ in random trees. These couplings are drawn from discrete multimodal distribution:
\begin{equation}
    \label{}
    \rho(C_{ij})=\sum_{l=1}^k\frac{a_l}{N}\delta(C_{ij}-c_l)+(1-\sum_{l=1}^k\frac{a_l}{N})\delta(C_{ij})
\end{equation}
 
At this step, the model can be considered as the union of $k$ disjoint random networks ("small world" random networks with finite connectivity \cite{nikoletopoulos2004replicated}). In the $l$-th network a given pair of clusters are connected with probability $a_l/N$  where the connectivity $a_l$ is the average number of connections per cluster (see Fig.~\ref{fig:Fig3} for illustrative explanation). The values $a_l$ remain finite in the thermodynamic limit $N \to \infty$. 
\begin{figure}
    \centering
    \includegraphics[width=\textwidth]{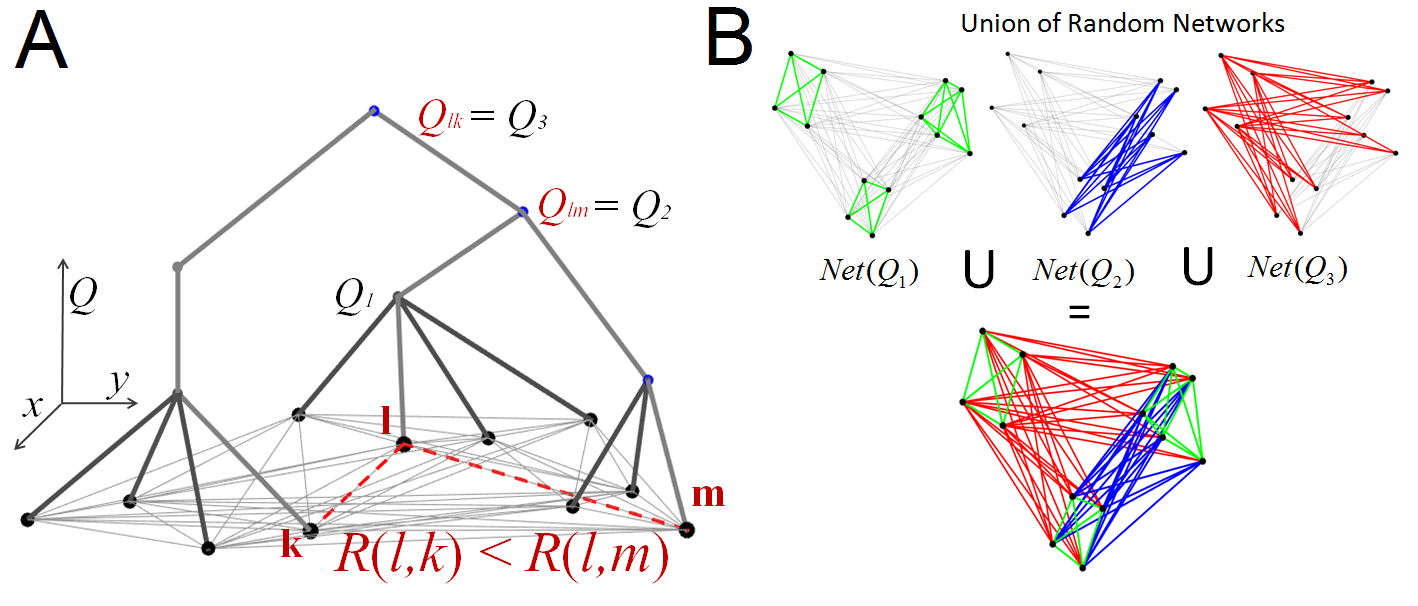}
    \caption{Schematic illustration of the different distances between the vertices of the graph - euclidean and ultrametric (A). From the graph one can see that the common path in tree for vertices $l$ and $m$ is greater than that for vertices $l$ and $k$. The correlation analysis shows that $U_l, U_m$ are strongly correlated and $U_l, U_k$ are almost non-correlated fields, despite the fact that the Euclidean distance $R(l,m)$ is greater than $R(l,k)$. This is contrary to the usual decay of correlations with distance in space. Schematic representation of the union of random networks with different connectivity (B).}
    \label{fig:Fig4}
\end{figure}

In the thermodynamic limit, the connectivity-averaged replicated partition function becomes to the leading order in $N$:
\begin{equation}
\label{eq:Xi_3}
Z(t,z)=\Big\langle\sum_{\phi_1...\phi_N}\prod_{i=1}^N\frac{1}{\phi_i!}\exp(-\beta H_0\sum_i\phi_i+\beta\sum_i\theta_i\phi_i+1/N\sum_{i\neq j}J(\phi_i,\phi_j)) \Big\rangle_{\{\theta_i\}}
\end{equation}
\noindent where we use the Hubbard-Stratonovich transform and introduce effective random fields $\theta_i$ with identical probability distribution $\rho(\theta)=\frac{1}{\sqrt{2\pi}\sigma}\exp\Big(-\frac{\theta^2}{2\sigma^2}\Big)$. Given pair of clusters $\phi_i, \phi_j$ interacts via pair potential of the form:
\begin{equation}
\label{eq:J_interaction}
    J(\phi_i,\phi_j)=\sum_{l=1}^ka_l(e^{\frac{\beta^2}{2}\phi_i\phi_jc_l}-1)
\end{equation}


\subsection{Analytical calculations: Replica Symmetric Ansatz}


Since the connectivities are finite, one cannot expand the inner exponential and introduce the order parameters, like in standard infinite-connectivity calculations \cite{sherrington1975solvable, schneider1977random, mezard1987spin}. Instead, to extract the variables into summation from the inner exponential, we use finite connectivity technique. Expression (\ref{eq:Xi_3}) for replicated partition function can be transformed into an integral to be calculated by steepest descent as $N\to\infty$, via the introduction of the order parameter distribution $P(\sigma)=1/N\sum_{i}\delta_{\sigma \phi_i}$ which represents the fraction of clusters $\phi_i$ of the size $\sigma$ (see Supplementary Materials for details):

\begin{equation}
\label{eq:Xi_4}
Z(t,z)=\int \prod_{\sigma}\Big[dP(\sigma)d\hat{P}(\sigma)\Big]e^{N\Psi(P,\hat{P})}
\end{equation}
\begin{equation}
\Psi(P,\hat{P})=i\sum_{\sigma}\hat{P}(\sigma)P(\sigma)+\sum_{\sigma\tau}P(\sigma)P(\tau)J(\sigma,\tau)+\log\sum_s\Big\langle\frac{1}{s!}\exp\Big[-\beta H_0s+\beta\theta s-i\hat{P}(s)\Big]\Big\rangle_{\{\theta\}} \nonumber
\end{equation}
\noindent where $\hat{P}$ is the auxiliary distribution. In the large-$N$ limit the integral is dominated by the stationary point of $\Psi(P,\hat{P})$. After eliminating the auxiliary order function  by means of saddle-point equations $\frac{\partial \Psi}{\partial P}=0$ and $\frac{\partial \Psi}{\partial \hat{P}}=0$, the self-consistency equation for density  has the form:
\begin{equation}
\label{eq:self-cons_eq}
 P(\sigma)=\frac{1}{\Gamma \sigma!}\Big\langle\exp\Big[ -\beta H_0 \sigma+\beta \theta \sigma +2\sum_\tau P(\tau)J(\sigma,\tau)\Big]\Big\rangle_{\{\theta\}}
\end{equation}
\noindent where $\Gamma$ remains a normalization factor, so $\sum_{\sigma} P(\sigma)=1$. Substituting saddle-point equations the potential $\Psi$ becomes
\begin{equation}
\label{eq:Psi}
\Psi(P)=-\sum_{\sigma\tau}P(\sigma)P(\tau)J(\sigma,\tau)+\log\sum_s\Big\langle\frac{1}{s!}\exp\Big[-\beta H_0s+\beta\theta s+2\sum_\tau P(\tau)J(s,\tau)\Big]\Big\rangle_{\{\theta\}}
\end{equation}
where $\Psi(P)$ has energetic and entropic contributions, $\Psi=\Psi_{en}+\Psi_{entr}$, that are, respectively, the first and the second terms of the right-hand side of (\ref{eq:Psi}). 
Equation (\ref{eq:self-cons_eq}) is a nonlinear functional equation on a discrete space of possible sigma values. An analytical solution to this equation is often not possible to obtain in a general form.
Equations of this kind arise in the theory of quenched disorder systems, especially in the theory of spin systems on random lattices and graphs. The essence of the solution method is similar to the classical mean field approach \cite{mezard1987spin}. However, instead of a constant external field, a certain ensemble of independent random fields at each site is introduced. So the self-consistent equation is formulated not for the value of the constant mean field, but for the distribution function of such random fields from ensemble. 


In work \cite{nikoletopoulos2004replicated} the system of spins interacting via one small world random network is considered using replicated transfer matrix technique. Despite significant differences some steps of our calculations are based on the model presented in this paper \cite{nikoletopoulos2004replicated}. To simplify the mathematical calculations, we leave only the principal (leading) term in the expression for the pair potential  (\ref{eq:J_interaction}) $J(\sigma, \tau)=c(e^{\beta\alpha\sigma\tau}-1)$, where $\alpha=\beta c_{max}/2$ and $c_{max}$ is the maximum value over  $ c_l$ in equation (\ref{eq:J_interaction}). We will look for the solution $P(\sigma)$ in the so-called Replica Symmetric (RS) ansatz \cite{nikoletopoulos2004replicated,erichsen2011phase, erichsen2017multicritical}. The ergodic, or RS ansatz corresponds to the distribution $P(\sigma)$ in the following form:

\begin{equation}
\label{eq:P_RS}
P(\sigma)=\frac{1}{\sigma!}\int dh W(h) e^{\beta \sigma h}/\chi_0(h)
\end{equation}

\noindent where $W(h)$ is the normalized distribution of local complex fields $h=h_1+ih_2$ ($dh=dh_1dh_2$), and $\chi_0(h)$ is an auxiliary partition function
\begin{equation}
\chi_\tau(h)=\sum_s \frac{1}{s!}e^{\beta s (h+\alpha\tau)}
\end{equation}

Introducing RS ansatz (\ref{eq:P_RS}) into equation (\ref{eq:self-cons_eq}) one can obtain the following self-consistent equation for local fields distribution $W(h)$ (see Supplementary Materials for details):
\begin{equation}
\label{eq:W_sol}
W(h)=\chi_0(h)e^{-c}\sum_{k\geq0}\frac{c^k}{k!}\Big\langle \int\prod_{l=1}^kdh_lW(h_l)\delta\Big(h+H_0-\theta-\frac{\alpha}{\beta}\sum_{l=1}^k \frac{\partial }{\partial h} \log \chi_0(h_l) \Big) \Big\rangle_{\theta}
\end{equation}
Numerical algorithm for solving this self-consistent equation for local fields distribution is presented below.


Using RS representation (\ref{eq:P_RS}) one can obtain from (\ref{eq:Psi}) energetic and entropic contributions $\Psi=\Psi_{en}+\Psi_{entr}$ 
\begin{eqnarray}
\Psi_{en}=-\sum_{\sigma, \tau}P(\sigma)P(\tau) J(\sigma, \tau)= \\
=-\int\int dh_1 dh_2 \frac{W(h_1) W(h_2) }{\chi_0(h_1) \chi_0(h_2)}\sum_{\sigma \tau}\frac{\exp(\beta h_1\sigma+\beta h_2 \tau)}{\sigma! \tau!}\sum_{l=1}^k a_l (e^{\beta \sigma \tau \alpha_l}-1) \nonumber
\end{eqnarray}
after linearization one can obtain the following expression:
\begin{equation}
\label{eq:Psi_en}
    \Psi_{en}(P)=-\frac{1}{\beta}\sum_{l=1}^ka_l\alpha_l \Big(\int dh W(h)\frac{\partial}{\partial h}\log \chi_0(h)\Big)^2
\end{equation}
The last term of $\Psi$ in (\ref{eq:Psi}) corresponds to the entropy contribution $\Psi_{entr}(P)$. By substituting the RS solution it becomes:
\begin{equation}
\label{eq:Psi_entr}
\Psi_{entr}(P)=\log \Big\langle \exp\exp \Big(-\beta U_0+\beta \theta +2\sum_{l=1}^ka_l\alpha_l\int dh W(h) \frac{\partial }{\partial h} \log \chi_0(h)\Big)\Big\rangle_{\{\theta\}}
\end{equation}
We observe that the distribution $W(h)$ can be represented by the shifting $W(h)=W_0(h-\frac{1}{\beta}\log\frac{t}{z})$, where $W_0$ is the solution of self-consistent equation with $t/z=1$. 
In the thermodynamic limit we are interested in the case $z\sim N$, so expressions (\ref{eq:Psi_en}) and (\ref{eq:Psi_entr}) can be approximated by the following leading terms:
\begin{equation}
\label{eq:Psi_en_leading}
\Psi_{en}(P)=-\frac{t^2}{z^2\beta}\sum_{l=1}^ka_l\alpha_l \Big(\int dh W_0(h)\frac{\partial}{\partial h}\log \chi_0(h)\Big)^2
\end{equation}

\begin{equation}
\label{eq:Psi_entr_leading}
\Psi_{entr}=\frac{t}{z}Z_0e^{\frac{\beta^2\sigma^2}{2}}+2\frac{t^2}{z^2}Z_0e^{\frac{\beta^2\sigma^2}{2}}\sum_{l=1}^ka_l\alpha_l \int dh W_0(h)\frac{\partial}{\partial h}\log \chi_0(h)
\end{equation}
Where $Z_0=e^{-\beta U_0}$. As results after substitution (\ref{eq:Psi_en_leading}) and (\ref{eq:Psi_entr_leading}) to (\ref{eq:F_Z-transform}) one can obtain desired average free energy for the first order expansion in fields correlation. Following the Free Energy Averaging technique \cite{forte2014effective, khlyupin2017random}, we equate the obtained free energy with the energy of the reference system in the effective field $U_{eff}$:
\begin{equation}
\label{eq:U_theory}
    U_{eff}(\beta)=U_0+U_{fluct}+U_{topology}=U_0-\frac{\beta \sigma^2}{2}+\frac{\bar{\alpha}}{\beta}\Big(2 \frac{m_0}{Z_0} e^{-\frac{\beta^2\sigma^2}{2}}-\frac{1}{\beta}\frac{m_0^2}{Z_0^2}e^{-\beta^2\sigma^2} \Big)
\end{equation}
\begin{equation}
    m_0=\int dh W_0(h)\frac{\partial }{\partial h}\log \chi_0(h) \nonumber
\end{equation}
\noindent where $\bar{\alpha}=\sum a_l\alpha_l/\sum a_l$ is averaged value of $\alpha$ via knowledge of random graph generation pattern. Thus the effective external potential is explicit function of the density distibution $W(h)$ which can be obtained numerically using self-consistent equation (\ref{eq:W_sol}) for certain temperature $\beta$ and the random branching graph statistics. For qualitative analysis $U_{topology}$ may be approximately simplified for the case of $U_0=0$ with $Z_0= exp{(-\beta U_0)}=1$:
\begin{equation}
\label{eq:U_simple}
    U_{topology}=\frac{\bar{\alpha}}{\beta}e^{-\frac{\beta^2\sigma^2}{2}}
    \int dh W_0(h)\frac{\partial }{\partial h}\log \chi_0(h) 
\end{equation}

\section{Results and Discussion}

\subsection{Numerical solution of self-consistent equation}

\begin{figure}
    \centering
    \includegraphics[width=\textwidth]{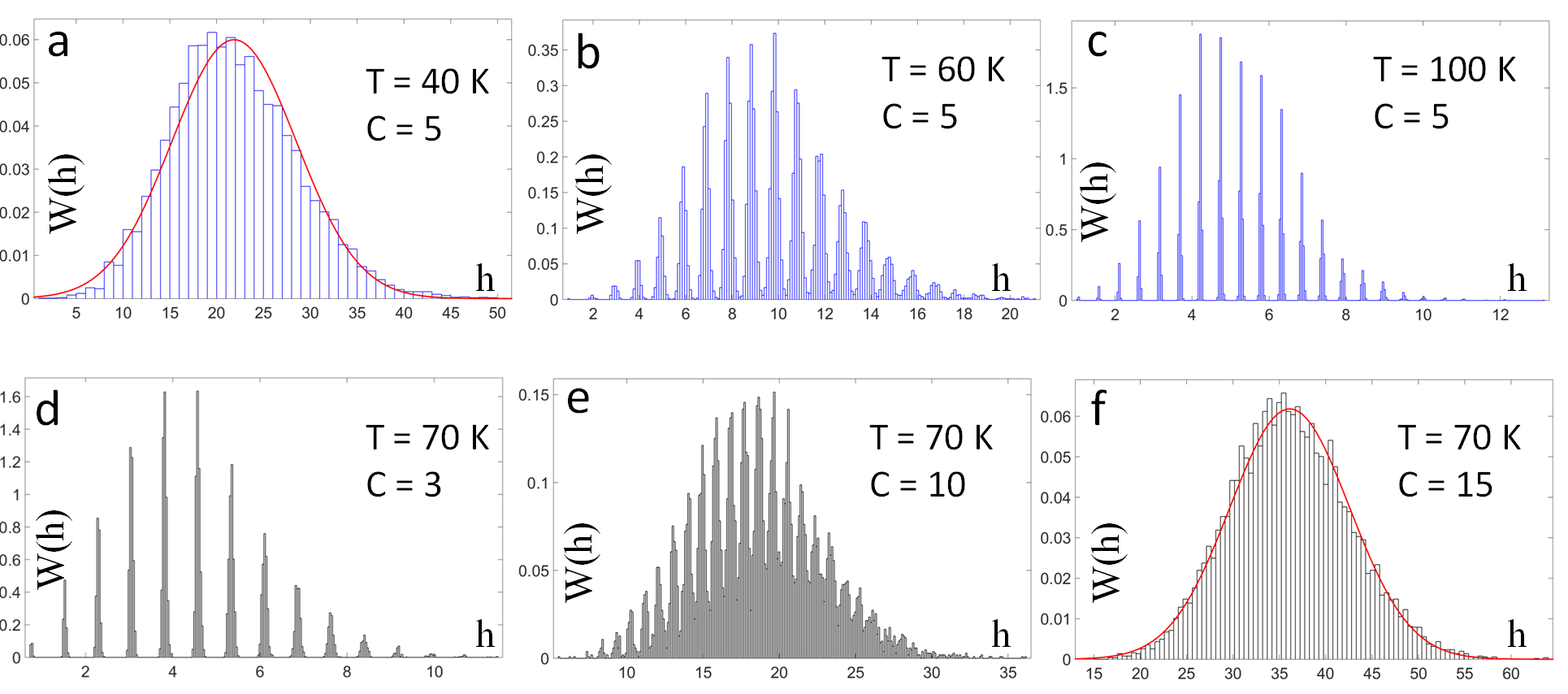}
    \caption{Different solutions of the integral equation for the distribution function of local fields in the absence of an external field $H_0=0$ and fluctuations $\theta=0$. Local fields distributions are presented for a fixed coupling constant and different temperatures (a, b, c) and for a fixed temperature and various coupling constants (d, e, f).}
    \label{fig:Fig5}
\end{figure}

\begin{figure}
    \centering
    \includegraphics[width=\textwidth]{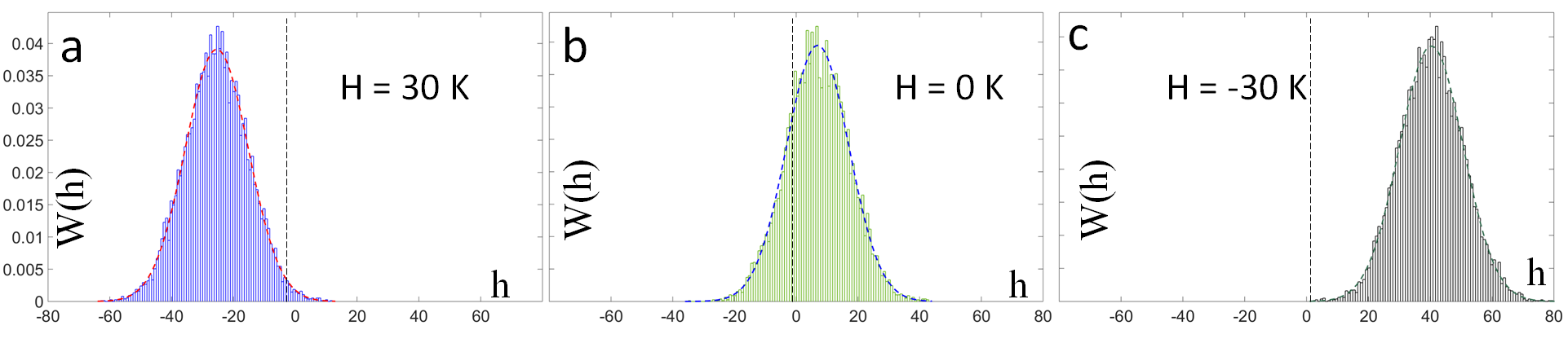}
    \caption{Different solutions of the integral equation for the distribution function of local fields for $\theta \neq 0$
    . Calculations were performed for the same coupling constants and temperatures (T = 80 K, C = 5) and different values of the external field (a, b, c).}
    \label{fig:Fig6}
\end{figure}

To solve the self-consistent integral equation (\ref{eq:W_sol}) for local fields density $W(h)$ we have proceeded numerically by means of population dynamics with large number of fields updated by an iterative method \cite{mezard2001bethe, erichsen2011phase}. 
Several factors affect the solution: temperature $T$, mean connectivity $c$, external field $H_0$ and random fluctuations $\theta$.  Thus, to investigate the behavior of the solution, we performed a series of calculations fixing some parameters and varying others. The results of numerical calculations are presented in Fig.~\ref{fig:Fig5} and Fig.~\ref{fig:Fig6}. To study the dependence on the temperature and the constants of the connectivity, we consider the case when the external field and the fluctuations are turned off. From the results it follows that the distribution is sensitive to both temperature and topological features of the graph (see Fig.~\ref{fig:Fig5}). With temperature increasing and small connectivity, the distribution becomes narrow and demonstrates multimodal structure, turning into a discrete spectrum. On the contrary, with temperature decreasing and increasing of connectivity, the distribution covers a wider range of field values and becomes smooth, approaching the normal distribution (see Fig.~\ref{fig:Fig5}a,f)
In the second series of calculations, we have fixed temperature $T = 80$ K, connectivity $c = 5$ and included random fluctuations $\theta$ with standard deviation $\sigma=10$ K and zero mean. In this case, the discrete distribution structure disappears. When the external field is switched on, a distribution shift is observed depending on the values of the external field, and the shape of the distribution varies slightly (see Fig.~\ref{fig:Fig6}).

\subsection{Comparison with Monte-Carlo simulation}
To validate the final theoretical predictions of the effective potential, we have performed a series of numerical experiments based on Monte Carlo simulation of random branching graphs. The following algorithm was used to generate a graph ensemble: starting from the root of the tree, for each parent node, number of child nodes was created according to the Poisson distribution with a mean value determined by a specific generating pattern. Each node of the graph contributes to the field according to the normal distribution with zero mean and variance $\Delta$: $\mathbb{E}\xi^2=\Delta^2$. This process continues until the required depth of the graph is reached. Recall that a random external field is created in the process of summing random variables along the branches of the graph from root to leaf. Then, for each graph realization, free energy has been directly calculated and global averaging has been performed over the entire ensemble according to equation (\ref{eq:LogF}). 
\begin{figure}[b]
    \centering
    \includegraphics[width=0.8\textwidth]{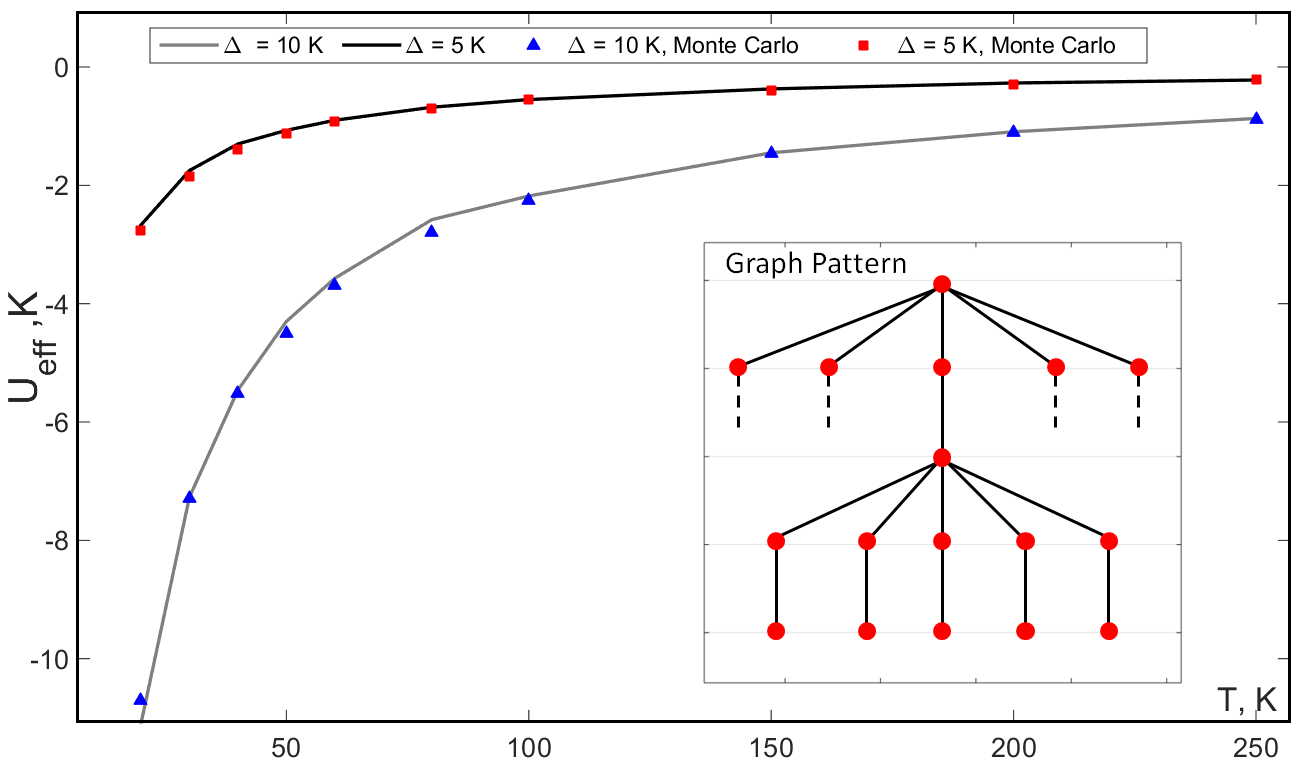}
    \caption{The temperature dependence of the effective potential is presented for different variances of random variables at vertices of random graphs $\mathbb{E}\xi^2=\Delta^2$ for $\Delta_1=5$ K and $\Delta_2=10$ K. The external field is turned off. The inset schematically shows the structural generation pattern of the graph ensemble.}
    \label{fig:Fig7}
\end{figure}

\begin{figure}
    \centering
    \includegraphics[width=\textwidth]{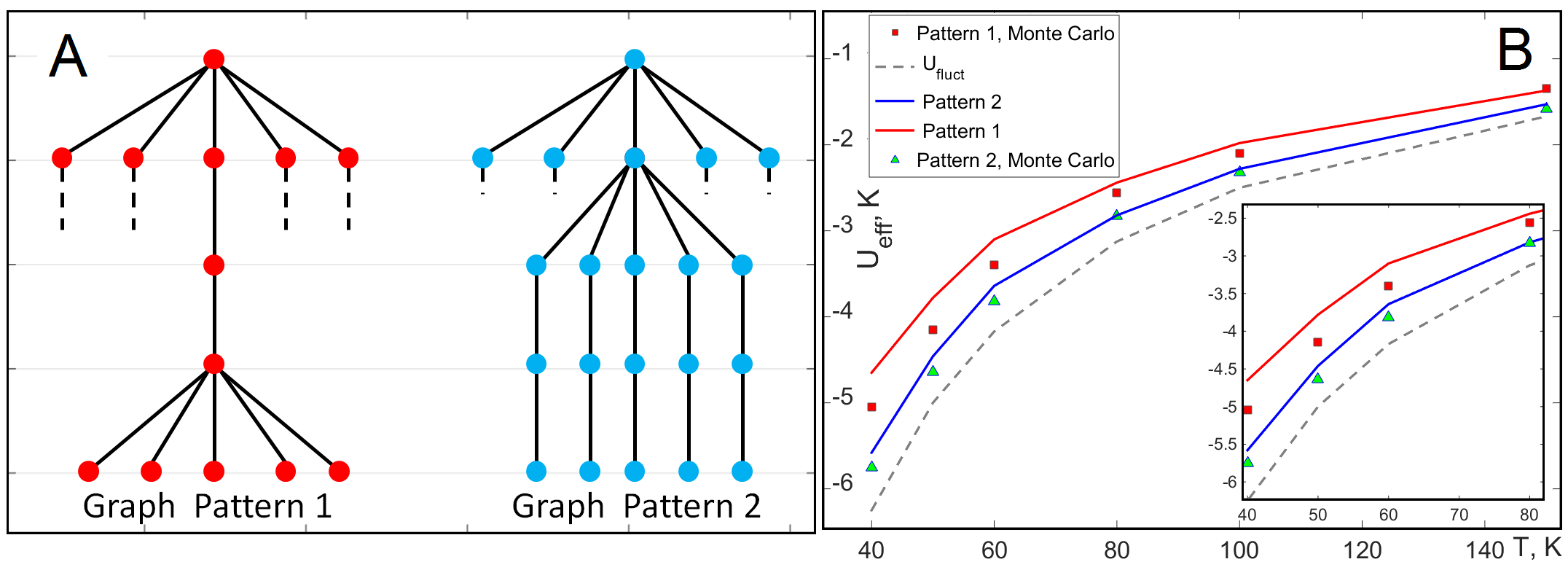}
    \caption{The dependence of the effective potential on temperature is presented for graphs with different structural patterns. The external field is turned off. Schematic representations of patterns (A). Temperature dependence of the effective potential for given generation patterns at $\Delta = 10$ K and comparison with Monte Carlo simulations (B). The dashed line indicates only the contribution from the $U_{fluct}$ term.}
    \label{fig:Fig8}
\end{figure}
In the current work, the size of the ensemble is $N=10^5$. Results of the comparison between theoretical calculations according to (\ref{eq:U_theory}) with Monte Carlo simulation carried out on an ensemble of graphs with different generating patterns are shown in Fig.~\ref{fig:Fig7} and Fig.~\ref{fig:Fig8}. In the first series of experiments, we have  investigated the effect of random field fluctuations on the effective potential. Based on the same generating pattern, cases with different values of the standard deviation of the fields at the nodes of the graph have been compared ($\Delta=5$ K and $\Delta=10$ K). The results demonstrate a good agreement between the theory and the numerical experiment. Fig.~\ref{fig:Fig7} shows that the temperature dependence is particularly strong at low temperatures and large fluctuations of random fields. Of greatest interest is the influence of the topology of random graphs on the free energy averaged potential. Therefore, in the second series of experiments, the simulation was carried out for two different generating patterns (see Fig.~\ref{fig:Fig8}) with equal standard deviation of local fields fluctuations at the nodes $\Delta=10$ K. For both patterns, the fluctuation part of the potential $U_{fluct}(T)$ is the same (depicted by a gray dashed line) and by itself cannot predict the differences observed in the simulation. However, the results of the theory demonstrate good agreement with experiment by taking into account the last term in the expression for the effective potential $U_{topology}(T)$.

\subsection{Realistic rough surface model}

In addition to highly perturbed hierarchical surfaces, molecular scale roughness is also of interest in many areas of practice. In such surfaces, the root-mean-square deviations of the surface height along the normal are of the order of several molecular sizes. 
In the case of a not highly developed hierarchical structure of the surface, the third term in the expression for the effective potential \eqref{eq:U_theory} can be neglected (from the previous analysis of the solution on Fig.~\ref{fig:Fig8}). This opens up scope for numerical analysis of the impact form first two terms without involving calculations using graph statistics.

\begin{figure}
    \centering
    \includegraphics[width=\textwidth]{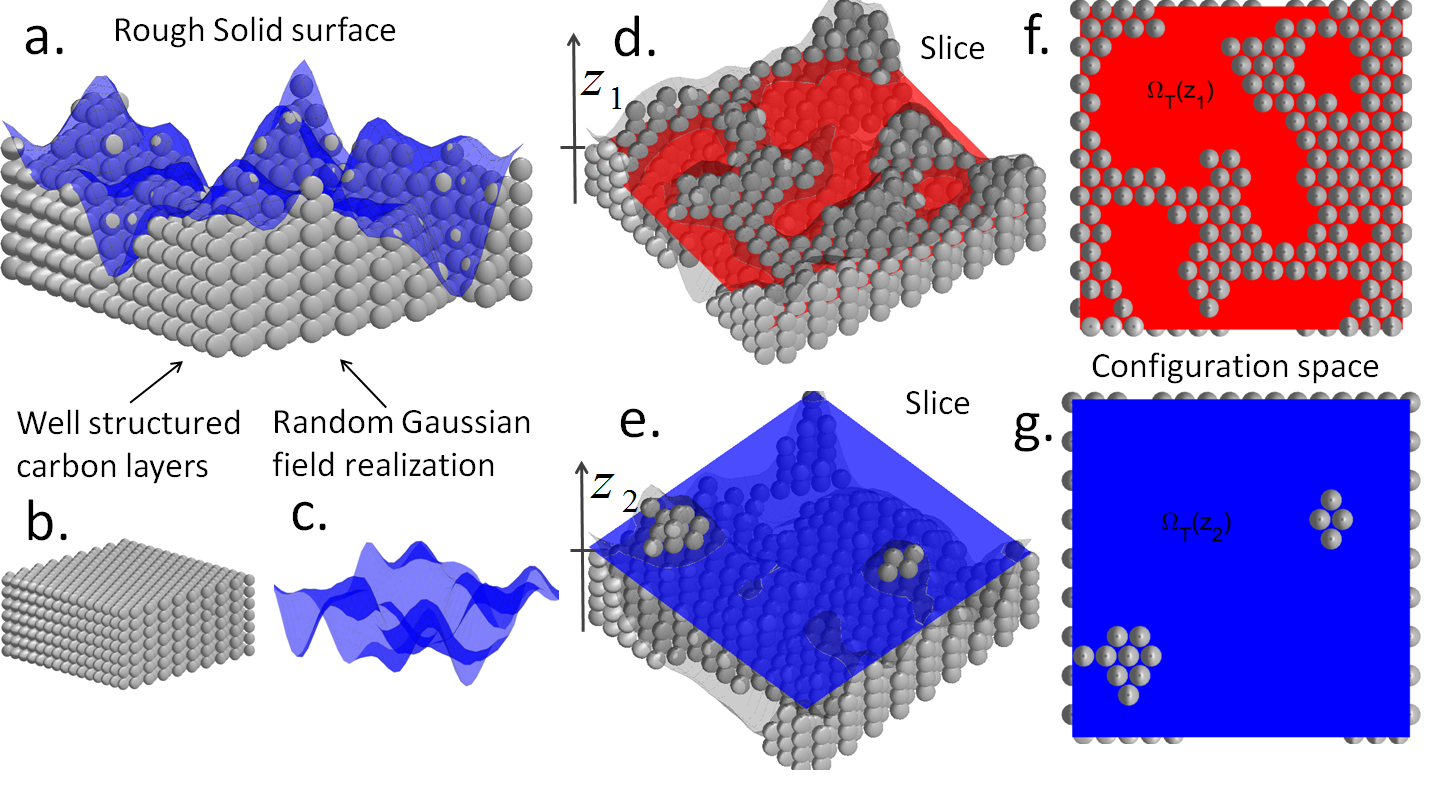}
    \caption{Illustration of 3D rough solid surface molecular model geometry (a). Gray balls illustrate solid lattice of carbon atoms (b). Solid surface coverage can be described by certain realization of the 2D random field (c). Configuration space $\Omega_T(z_1)$ available for fluid molecule at $z_1 = -1.0~D$: side view (d) and view from above (f). Configuration space $\Omega_T(z_2)$ available for fluid molecule at $z_2 = 1.5~D$: side view (e) and view from above (g), where $D$ - carbon atom diameter.}
    \label{fig:Fig9}
\end{figure}

An ensemble of realizations of rough solid surfaces with given average geometric properties is artificially generated using Monte Carlo simulations. Further, for each such surface model from the ensemble, the configuration space available for the fluid molecule $\Omega_T(z)$ can be extracted  with a certain step along the normal direction $z$. Then $x,y$-coordinates of the fluid molecule are chosen randomly in $\Omega_T(z)$ and for each such point the sum of all pair interactions with solid atoms is calculated numerically. Thus, for each slice $z$, one have a data sample of the interaction energies for various possible realizations of configuration spaces within the ensemble of solid surfaces. Given the sample data, it is possible to statistically estimate the means $U_0(z)$ for each $z$ and the standard deviations, which correspond to the $\sigma(z)$ from the (\ref{eq:U_theory}). The characteristic behavior of these quantities depending on the distance from the surface $z$ determines the temperature effects on effective fluid-solid potential within the given approximation. For these purposes, we have implemented a numerical algorithm in the MATLAB development environment. To calculate the potential, a molecular model of a rough solid surface is generated as follows. First, a 3D solid model is created from several structured 2D layers of carbon atoms (in this implementation, 8 layers are used, see Fig.~\ref{fig:Fig9}.b). Then realization of a 2D Gaussian field $Z(r)$ with given statistical geometric properties is randomly generated (see Fig.~\ref{fig:Fig9}.c). This field is used as a coating on top of a solid surface, namely, carbon atoms lying above $Z(r)$ are removed from the 3D model. Thus, the Figure~\ref{fig:Fig9}.a shows obtained solid surface model. 

In this work we use correlated Gaussian random field as one of the modern basic models describing real corrugated solid geometry. In the frame of this approach random rough surface can be characterized by two natural parameters: $\delta$ corresponds to variance of the fluctuating surface height $Z(r)$ along the normal direction and correlation length $\tau$ reflects the scale of surface fluctuations in the lateral plane (for further reading, see our work \cite{aslyamov2017density}).  

\begin{figure}
    \centering
    \includegraphics[width=\textwidth]{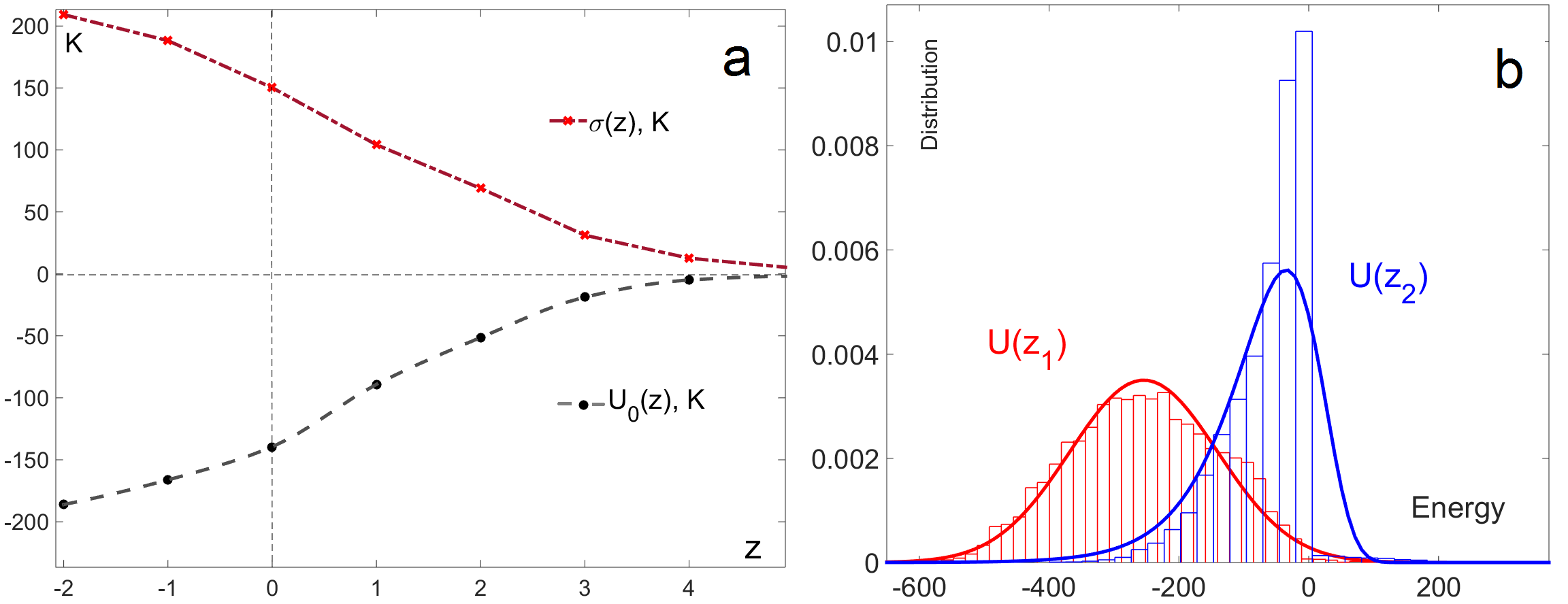}
    \caption{Values of $U_0(z)$, $\sigma(z)$ calculated with a discrete steps along $z$, the x-axis in reduced units with respect to carbon atom diameter D (a). Distribution of the interaction energy values $U(z_1)$ (red line) in $\Omega_T(z_1)$ and $U(z_2)$ (blue line) in $\Omega_T(z_2)$ (b).} 
    \label{fig:Fig10}
\end{figure}
For a specific numerical analysis in this work, we consider the correlated Gaussian random field as the model of the rough solid with the following geometrical parameters $\delta=1.5 D$,  $\tau=2.0 D$,  where $D=3.4~\AA$ is the carbon atom diameter (hereinafter we  use reduced units with respect to diameter of carbon atom). Thus, in the calculation process, an ensemble of $N = 10^3$ surfaces with such parameters is realized (an example of one implementation is just shown in Figure~\ref{fig:Fig9}). The x,y-size of the full molecular models is 40 $D$, however, for calculating $\Omega_T(z)$, the central region of size 20 $D$ is taken to eliminate edge effects in the process of the interaction energy calculation. Argon with the potential of carbon-argon intermolecular interaction in the form of Lennard-Jones was taken as a fluid molecule:
\begin{equation}
    U_{LJ}=4\varepsilon_{sf}\Big[\left(\frac{\sigma_{sf}}{r}\right)^{12}- \left(\frac{\sigma_{sf}}{r}\right)^6\Big]
\end{equation}
where parameters of solid-fluid interaction $\varepsilon_{sf}, \sigma_{sf}$  were calculated according to the standard Lorentz-Berthelot chemical mixing rule: $\varepsilon_{sf} = \sqrt{\varepsilon_{ss}\varepsilon_{ff}}, \quad \sigma_{sf} = \dfrac{1}{2}\left(\sigma_{ss} + \sigma_{ff}\right)$ with fluid-fluid and solid-solid parameters \cite{neimark2009quenched}: $\varepsilon_{ss} = 28$~K, $\sigma_{ss} = 3.4~\AA, \varepsilon_{ff} = 111.95$~K, $\sigma_{ff} = 3.358~\AA$. 

\begin{figure}
    \centering
    \includegraphics[width=0.8\textwidth]{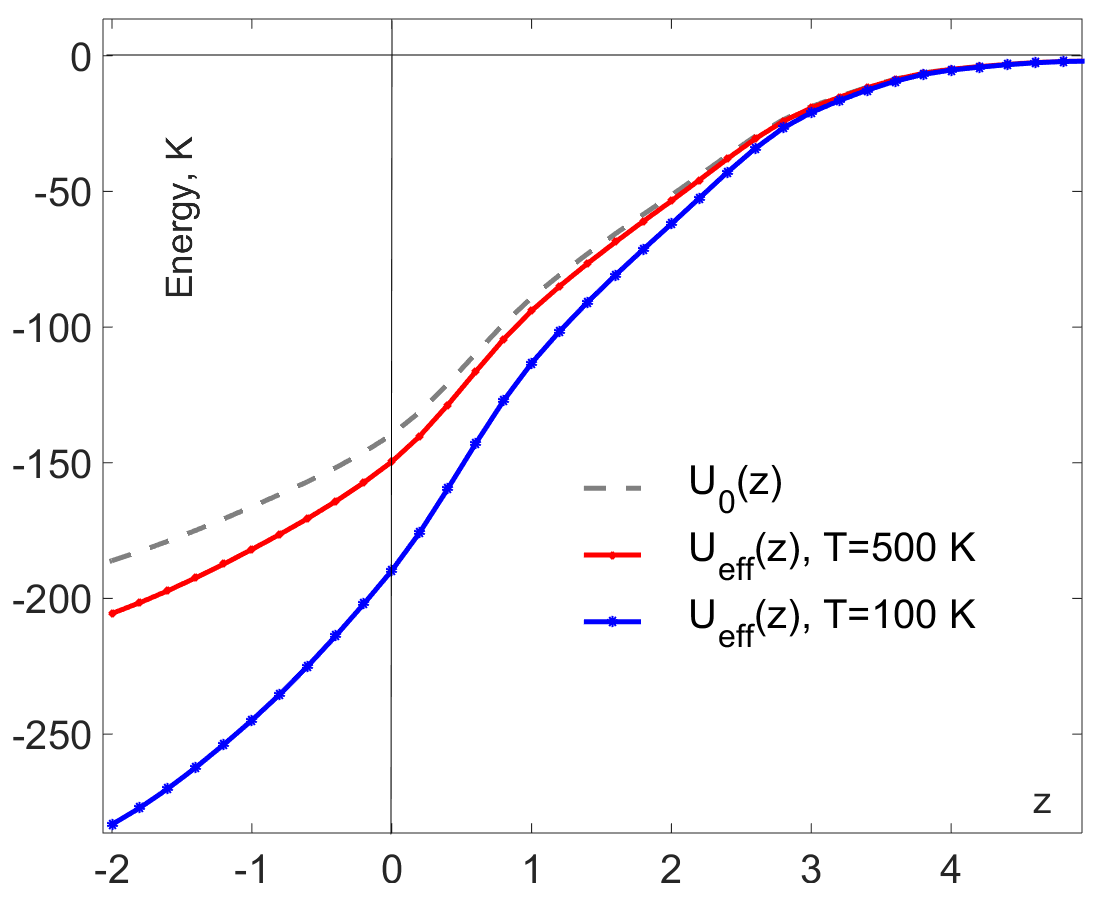}
    \caption{The behaviour of the effective potential $U_{eff}(z)$ for two different temperatures $T_1=500 K$ (red solid line), $T_2=100 K$ (blue solid line). The dashed line corresponds to $U_0(z)$} 
    \label{fig:Fig11}
\end{figure}
Thus, collecting the above together, one can numerically analyze functions $U_0(z)$, $\sigma(z)$ depending on the z coordinate along the normal.The necessary calculations were performed in the range from -2$D$ to 5$D$ along $z$. To study the characteristic behavior of fluctuations of interaction energy within the $\Omega_T(z)$, consider two characteristic slices $\Omega_T(z_1)$ at the level $z_1=-1.0~D$ and  $\Omega_T(z_2)$ at the level $z_2=1.5~D$ (see Fig.~\ref{fig:Fig9} d,f and  Fig.~\ref{fig:Fig9} e,g respectively). Figure~\ref{fig:Fig10}.B  shows histograms of the interaction energy values in these layers, taking into account the modeling over the entire ensemble of surface realizations. Naturally, the average energy value is less when the fluid molecule moves away from the surface ($z_2>z_1$). Of greater interest are the standard deviations reflecting the scale of fluctuations. It can be noted that the more developed, inhomogeneous landscape of the surface (and, accordingly, the structure of the configuration space) the greater the dispersion and spread of values is demonstrated by the energy sampling within the slice. It is also worth noting that the deeper the molecule penetrates into the solid, the more accurately the distribution is described by the Gaussian model (distribution of $U(z_1)$). And vice versa, with distance from the solid surface, the distribution loses its Gaussian shape and becomes more asymmetric (distribution of $U(z_2)$). However, in this region, the effective potential weakly depends on temperature due to the small scale of fluctuations. This observation confirms the theoretical assumptions of the approach developed above, where a multivariate Gaussian distribution is used as a probabilistic model of random fields deep in solid. These properties are reflected in the behavior of the effective potential $U_{eff}(z;\beta)$ as a function of $z$ for different temperatures. Figure~\ref{fig:Fig11}  shows the behavior of the potential obtained by equation (\ref{eq:U_theory}) based on calculated values $U_0(z)$, $\sigma(z)$ (see Fig.~\ref{fig:Fig10}.A). It can be seen that the effect of temperature is different depending on $z$ and the greater the greater the standard deviation $\sigma(z)$.

\section{\label{sec:conclusion} Conclusions and Discussion}

We described a fluid molecule's thermal properties near solid surfaces with nanoscale geometric heterogeneity at a wide range of temperatures and heterogeneity scales. Our model shows excellent agreement with the Monte Carlo simulation for several cases with different graph patterns reflecting the solid geometry's hierarchical properties. Moreover, we found that at low temperatures, a significant contribution to free energy comes from random geometry's hierarchical structure. For example, two solid surfaces with the same roughness affect the fluid molecule differently due to their graph representation's different topology. Thus, the proposed model looks promising for the further construction of effective fluid-solid interaction potentials. These potentials are highly desirable in Molecular Dynamics or DFT modeling of a broad class of interface thermodynamic properties. In a recent series of works, DFT-based models have been developed for the fluid interaction with geometrically heterogeneous surfaces. More specifically in \cite{aslyamov2017density, aslyamov2019random} we proposed novel Random Surface Density Functional Theory (RS-DFT) and Random Surface Statistical Associating Fluid Theory (RS-SAFT) to describe the adsorption of simple and chain fluids on various modern carbon materials with nanoscale heterogeneity, respectively. These approaches are based on the theoretical model of the effective fluid-solid potential but without proper temperature dependence.  Current research results expand the boundaries of its applicability to low temperatures and large geometric defects and constitute a significant step in understanding the influence of random geometry on fluid thermodynamics. 

The statistical model on random graphs and theoretical methods proposed in this paper may be interesting for studying cooperative phenomena in biological populations and social networks. The application of statistical physics methods in the study of complex systems dynamics, social behavior, and cooperation in populations and swarms is a hot topic of research \cite{forrow2018functional,adami2018thermodynamics,korolev2015evolution,filella2018model, barberis2016large}. For example, in work \cite{adami2018thermodynamics}, the authors study the "thermodynamics" of evolutionary games to answer how cooperation can evolve between players. They use the Hamiltonian dynamics of Ising type models to describe populations of cooperating and defecting players. The authors show that cooperators' equilibrium fraction is given by the expectation value of a thermal observable akin to a magnetization. Statistical models on graphs have modern applications in Computer Vision and Bayesian Networks to reconstruct images with defects \cite{yasuda2015statistical}, analysis of computational complexity and neural networks \cite{wemmenhove2003finite,monasson1999determining}. 
However, established models work well for pairwise interactions depending on the Euclidean distance in space \cite{parisi2002euclid}, while social interactions in a network or a swarm are much more complicated. In our proposed model, the "distance" is defined in terms of a random graph's topological structure, which opens up new possibilities for research applications.

\bibliography{apssamp}
\bibliographystyle{unsrt}
\end{document}